# The elephant in the room: The problem of quantifying productivity in evaluative scientometrics

Ludo Waltman, Nees Jan van Eck, Martijn Visser, and Paul Wouters

Centre for Science and Technology Studies, Leiden University, The Netherlands
{waltmanlr, ecknjpvan, visser, p.f.wouters}@cwts.leidenuniv.nl

In a critical and provocative paper, Abramo and D'Angelo claim that commonly used scientometric indicators such as the mean normalized citation score (MNCS) are completely inappropriate as indicators of scientific performance. Abramo and D'Angelo argue that scientific performance should be quantified using indicators that take into account the productivity of a research unit. We provide a response to Abramo and D'Angelo, indicating where we believe they raise important issues, but also pointing out where we believe their claims to be too extreme.

## 1. Introduction

In a critical and provocative paper, Abramo and D'Angelo (in press; hereafter AA) claim that commonly used scientometric indicators such as the mean normalized citation score (MNCS; Waltman et al., 2011) are completely inappropriate as indicators of scientific performance. AA argue that scientific performance should be quantified using indicators that take into account the productivity of a research unit (e.g., an individual researcher, a research group, a research institution, or an entire country). An example of such an indicator is the fractional scientific strength (FSS). This indicator is used extensively by AA in scientometric analyses in Italy. An in-depth discussion of the FSS is provided in an earlier paper by AA (Abramo & D'Angelo, 2014).

Below we provide a response to AA, indicating where we believe they raise important issues, but also pointing out where we believe their claims to be too extreme.

## 2. The problem of quantifying productivity

The key element in the criticism of AA is that commonly used scientometric indicators of scientific performance do not take into account the productivity of a research unit. Indicators such as the MNCS are obtained by calculating the total field-normalized number of citations of the publications of a research unit and by dividing this number by the number of publications of the research unit. These indicators provide a proxy of the average scientific impact of the publications of a research unit, but they do not take into consideration the productivity of the research unit. An indicator of productivity can be obtained by dividing the number of publications of a research unit by the number of researchers, or alternatively, by the amount of money spent on research. A next step then could be to construct a combined indicator of impact and productivity, which can be done by multiplying a productivity indicator and an impact indicator. Such an indicator can for instance be calculated by dividing the field-normalized number of citations of a research unit by the number of researchers, and in essence this is what is done by the FSS that is advocated by AA.[1]

Why is productivity not taken into account in commonly used scientometric indicators? There are at least two reasons for this. The first reason is that quantifying productivity requires not only data on scientific outputs (e.g., publications and their citations) but also data on scientific inputs (e.g., researchers or research funding). This input data often is not available, or it is not of sufficient quality. For instance, the 2015 edition of our CWTS Leiden Ranking includes 750 universities from 48 different countries. It is not feasible to obtain accurate and comparable input data for these 750 universities, and therefore our ranking is necessarily restricted to indicators that are based on output data only. The second reason is that scientometric analyses often aim to provide statistics that have been corrected for differences among scientific fields in publication and citation practices and that have been standardized with respect to an international baseline. In the case of indicators of productivity, obtaining statistics that meet these criteria is extremely challenging. Essentially, it requires the availability of input data not only for the research units for which indicators need to be calculated but for all research units worldwide. In addition, as is also pointed out by AA, input data needs to be available at a high level of detail. For instance, knowing the total number of researchers or the total amount of research funding of a university is not sufficient. Instead, input data needs to be available at the level of individual scientific fields. Moreover, even if we know the number of researchers of a university in a given field, we still do not have sufficiently detailed information. We also need to know how much time these researchers are able to spend on research and how much time they need to spend on other tasks, such as teaching. Furthermore, in the collection of input data, international standardization is required. There need to be international standards for determining the number of researchers or the amount of research funding of a research unit. Also, an internationally standardized classification system of scientific fields is needed.

---

[1] The terminology used by AA is somewhat different from ours. We distinguish between the concepts of productivity and impact. The concept of productivity used by AA combines our concepts of productivity and impact.

We fully agree with AA that productivity is a key element of scientific performance. The fact that in most evaluative scientometric analyses productivity is not taken into account is a big limitation, and it represents one of the most important shortcomings of evaluative scientometric analyses. Unfortunately, however, we are skeptical about the feasibility of scientometric measurements of productivity. The work done by AA in the Italian context is of great interest, and their ideas may be transferable to other countries with research systems that are organized in a relatively centralized way. However, at an international level, collecting standardized input data requires a high degree of coordination between countries, and it is probably not realistic to expect that this degree of coordination can be achieved.

## 3. Scientific performance as a multidimensional concept

AA claim that any indicator that does not capture both the productivity and the scientific impact of a research unit is not an indicator of scientific performance. We believe that this is based on a rather narrow concept of scientific performance, and we consider it more useful to adopt a pluralistic idea of scientific performance. In our perspective, scientific performance is a concept that has many different dimensions. Some dimensions of scientific performance can be quantified more easily than others, and different indicators are required for quantifying different dimensions of scientific performance.

To provide an illustration of the idea of scientific performance as a multidimensional concept, we consider two basic dimensions of scientific performance. These are the size-dependent and the size-independent dimension. The size-dependent concept of scientific performance is about the overall contribution of a research unit to science, irrespective of the size of the research unit in terms of the number of researchers or the amount of research funding. The size-independent concept of scientific performance is about the contribution of a research unit to science relative to the size of the unit. In this concept of scientific performance, a correction is made for differences between research units in their size. The size-independent concept of scientific performance is the concept that AA have in mind and that they seem to consider the only valid idea of scientific performance.

Suppose someone asks the following question to a scientometrician: Which universities in the Netherlands perform best in the field of physics? There is no unique scientometric answer to this question. For instance, a physicist looking for a position at a Dutch university may be interested to know which Dutch universities make the largest contribution to the field of physics. The physicist might expect that these universities will provide the most stimulating research environment. On the other hand, a policy maker may be interested to know which Dutch universities produce contributions to the field of physics in the most efficient way. The policy maker may consider these universities to be most successful in allocating the scarce resources available for scientific research. For the physicist, the most useful information is probably provided by a size-dependent indicator of scientific performance. Such an indicator could be obtained by calculating the number of publications, the number of citations, or the number of highly cited publications of the physics departments of the different Dutch universities. However, for the policy maker, the most useful information is provided by a size-independent indicator of scientific performance. For instance, if the FSS could be calculated for Dutch physics departments, this would probably offer helpful information to the policy maker.

The above example illustrates the need to distinguish between different dimensions of scientific performance. The size-dependent dimension is relatively easy to quantify. Size-dependent indicators of scientific performance aim to capture the production and impact of a research unit. These indicators can be based exclusively on output data. The size-independent dimension of scientific performance is much more difficult to quantify. Proper size-independent indicators of scientific performance, such as the FSS, aim to capture the production and impact of a research unit relative to the size of the research unit. These indicators require not only output data but also input data, for instance the number of researchers of a research unit or the amount of research funding. When no input data is available to quantify the size of a research unit, scientometricians typically use the number of publications of a research unit as a surrogate measure of the unit's size. This leads to indicators such as the MNCS, in which the field-normalized number of citations of a research unit is divided by the number of publications of the research unit. These indicators provide a proxy of the average impact of the publications of a research unit, but they completely disregard the productivity of a research unit. Consequently, these indicators capture the size-independent concept of scientific performance only in a partial way.

As already mentioned above, ignoring productivity is a big limitation. Does this mean that the MNCS and other similar types of size-independent indicators "are not worthy of further use or attention", as claimed by AA? We disagree with this strong claim of AA. We believe that the MNCS can still be a useful indicator, provided that users of the indicator are aware of the fact that it takes into account only impact and not productivity. In some cases, users may be willing to make the assumption that the effect of ignoring productivity is not too large and users may therefore accept the MNCS as a reasonable indicator of size-independent scientific performance, albeit an indicator with a limited precision. In other cases, for instance when the MNCS is used to support an expert committee responsible for the evaluation of a research unit, the committee members may use the MNCS to get insight into the average impact of the publications of the research unit, while they may use their own expert judgment to assess the productivity of the research unit.

From a certain perspective, the MNCS also has an advantage over indicators such as the FSS. The FSS takes into account only the most standard type of scientific output, namely publications in journals covered in bibliographic databases. Other types of scientific outputs, such as publications in non-covered journals, book publications, data sets, and software tools, are not considered in the FSS. Likewise, other ways of contributing to science, for instance by serving as editor of a journal, are not taken into account. Because non-standard scientific outputs are not considered in the FSS, the indicator will often provide an incomplete picture of the performance of a research unit. Moreover, if some research units put more effort into producing non-standard scientific outputs than other research units, there will be a bias in the FSS against the former units and in favor of the latter ones. The MNCS does not have this problem. Like the FSS, the MNCS takes into account only the most standard type of scientific output, that is, publications in journals covered in bibliographic databases. Other types of scientific outputs are not considered. However, unlike the FSS, the MNCS does not penalize research units for putting effort into producing non-standard scientific outputs. This is because the MNCS corrects for the size of a research unit not based on input data but based on the number of standard scientific outputs produced by the unit. Hence, in the MNCS, not only the 'output side' but also the 'input side' is determined based on standard scientific outputs, and this consistency between the input and output sides ensures that there is no penalty for putting effort into producing non-standard scientific outputs.

It is also possible to look at the difference between the FSS and the MNCS from a somewhat different perspective. The FSS promises to deliver what we ideally would like to have, namely a combined indicator of impact and productivity. However, because of the difficulties discussed above, it is not entirely clear to what degree the FSS is really able to deliver what it promises. On the other hand, the MNCS is more modest in what it promises to deliver, namely an indicator of impact only. Because it is more modest, the MNCS is able to deliver what it promises and therefore the MNCS avoids the risk of raising expectations that cannot be fulfilled.

## 4. How to move forward?

AA raise important issues that require serious attention from the scientometric community. However, there are no easy solutions. We believe that the community should take the following three steps to move forward:

1. *Create more awareness of the productivity problem.* AA deserve credit for drawing attention to the productivity problem. Scientometricians have put a lot of effort into improving the measurement of scientific impact (for a review of the literature, see Waltman, 2016), but they are often silent about the more difficult problem of quantifying productivity. University rankings, scientometric analysis tools such as InCites and SciVal, and many scientometric studies rely heavily on size-independent indicators such as the MNCS, but they typically do not explicitly draw attention to the fact that these indicators capture only impact and not productivity. Scientometricians should be more pro-active in this respect and should communicate the productivity problem in a more explicit way. At our center, we have occasionally used productivity indicators in our analyses (e.g., Moed, 2000) and we regularly discuss the problem of quantifying productivity in scientometric courses that we offer. However, we acknowledge that in other contexts, for instance in our CWTS Leiden Ranking, we need to find ways to more explicitly explain the productivity problem to the users of our indicators.

2. *Perform empirical analyses of the consequences of the productivity problem.* Scientometricians need to improve their understanding of the consequences of the productivity problem. This requires empirical analyses in which indicators such as the MNCS and the FSS are compared. AA present some first results of a comparative analysis, but more extensive analyses are needed. In their analysis, AA compare the MNCS and the FSS both at the level of individual researchers and at the level of universities. In our view, comparisons at the individual researcher level are less relevant. At this level, the use of the MNCS generally should not be recommended. A size-independent indicator such as the MNCS can be useful to compare research units that are of different size, but at the individual researcher level there essentially are no size differences. At other levels of aggregation, comparisons between indicators such as the MNCS and the FSS are highly relevant. If strong correlations are observed at these levels, this would suggest that the productivity problem has only limited consequences and this could justify the use of the MNCS as an indicator of size-independent scientific performance. On the other hand, if correlations turn out to be weak, the use of the MNCS would be more problematic. Ideally, indicators such as the MNCS and the FSS should be compared not only with each other but also with peer review outcomes.

3. *Explore the possibilities for more frequent use of input data in scientometric analyses.* According to AA, if governments and research institutions "expect ever more precise and reliable performance evaluations ... they must be prepared to give scientometricians the underlying data necessary for the job". In principle, we support this statement. However, as already emphasized above, the challenges involved in collecting standardized input data should not be underestimated. Collecting this data might be possible at a national level, but at an international level it probably will not be feasible, perhaps with an exception for countries that have partly integrated their research systems, such as the countries of the European Union. Despite these difficulties, we feel that scientometricians could pay more attention to the possibility of using input data in their analyses, and we believe that scientometricians should improve their knowledge of the current availability of input data. Moreover, scientometricians should investigate more deeply what types of input data are needed to construct meaningful productivity indicators, and they should explore possible ways of obtaining this data. This could perhaps be done in the context of the development of current research information systems (e.g., Sivertsen, 2016) and other new infrastructures for research data.

## 5. Conclusion

AA draw attention to the elephant in the room of evaluative scientometrics: Productivity is one of the key aspects of scientific performance, but it is not taken into account in commonly used scientometric indicators such as the MNCS. A number of statements made by AA are too extreme, in particular their complete rejection of any concept of scientific performance different from their own and also their call to journal editors, editorial board members, and referees to no longer accept publications dealing with 'invalid' indicators such as the MNCS. Nevertheless, the issues raised by AA are important and call for concrete action from the scientometric community. The community should create more awareness of the productivity problem and should study the consequences of the problem. Moreover, possibilities for more frequent use of input data should be explored.

## Acknowledgement

We would like to thank the members of the Advanced Bibliometric Methods working group at our center for helpful discussions.